\documentclass[useAMS]{mn2e}
\usepackage{times}
\input{epsf}

\title[RE J1034+396: The origin of the soft X-ray excess and QPO]
{RE J1034+396: The origin of the soft X-ray excess and QPO}

\author[M.Middleton, C. Done, M. Ward, M. Gierli{\'n}ski, N. Schurch]
{Matthew Middleton$^1$, Chris Done$^1$, Martin Ward$^1$, Marek Gierli{\'n}ski$^1$ and Nick Schurch$^1$\\
$^1$Department of Physics, University of Durham, South Road, Durham
DH1 3LE,
UK\\
}

\pagerange{\pageref{firstpage}--\pageref{lastpage}} \pubyear{2008}
\def\etal{et al.~\/}

\def\la{\mathrel{\hbox{\rlap{\hbox{\lower4pt\hbox{$\sim$}}}{\raise2pt\hbox{$<$}}
}}}
\def\ga{\mathrel{\hbox{\rlap{\hbox{\lower4pt\hbox{$\sim$}}}{\raise2pt\hbox{$>$}}
}}}
\def\d25{D$_{25}$}

\def\deg{\hbox{$^\circ$~\/}}

\begin{document}

\topmargin = -0.5cm

\maketitle

\label{firstpage}

\begin{abstract}

The X-ray quasi-periodic oscillation (QPO) seen in RE~J1034+396
is so far unique amongst AGN. Here we look at the another unique
feature of RE~J1034+396, namely its huge soft X-ray excess, to
see if this is related in any way to the detection of the QPO. We
show that all potential models considered for the soft energy
excess can fit the 0.3--10~keV X-ray spectrum, but that the
energy dependence of the rapid variability (which is dominated by
the QPO) strongly supports a spectral decomposition where the
soft excess is from low temperature Comptonization of the disc
emission and remains mostly constant, while the rapid variability
is produced by the power law tail changing in normalization. The
presence of the QPO in the tail rather than in the disc is a
common feature in black hole binaries, but low temperature
Comptonization of the disc spectrum is not generally seen in
these systems. The main exception to this is GRS~1915+105, the
only black hole binary which routinely shows super-Eddington
luminosities. We speculate that super-Eddington accretion rates
lead to a change in disc structure, and that this also triggers
the X-ray QPO.

\end{abstract}

\section{Introduction}

The discovery of a significant quasi-periodic oscillation in the
X-ray light curve of a Narrow Line Seyfert 1 AGN RE J1034+396
(Gierli{\'n}ski et al. 2008) strengthens arguments stressing the
similarities in the physics of the accretion flow between
supermassive and stellar mass black holes.  Previous evidence for
a simple correspondence between AGN and the black hole binaries
(BHB) included similarities in the broadband shape of the X-ray
variability power spectra, with characteristic break timescales
scaling with mass (e.g. M$^c$Hardy et al. 2006), but the
characteristic QPOs often seen in BHB light curves remained
undetected until now (Vaughan \& Uttley 2005; Leighly 2005).

The QPOs in BHB can be split into two main groups, at high and
low frequencies, respectively.  While there is as yet no clear
mechanism for producing either set of QPOs (or the correlated
broad band variability) there are many pointers to their origin
from the observations. Most clearly, their amplitude increases
with increasing energy. The BHB spectra typically contain two
components, a disc and tail, and this increasing amplitude is
consistent with the QPO (and all the rest of the rapid
variability) being associated with a variable tail while the disc
remains constant.  This shows that QPOs are produced by some mode
of the hot coronal flow rather than a mode of the thin disc. (see
e.g. the reviews by Van der Klis 2004; McClintock \& Remillard
2006; Done, Gierli{\'n}ski \& Kubota 2007).

The rest of the QPO properties are more complex.  The high
frequency (HF) QPO may have a constant fundamental frequency
(though it is seen at 3:2 harmonic ratios of this), which may
relate to the mass of the black hole. By contrast, the frequency
(and other properties) of the low frequency (LF) QPO change
dramatically with mass accretion rate, correlating with the
equally dramatic changes in the source spectra. Typically these
show that at low mass accretion rates compared to Eddington,
$L/L_{\rm Edd} \ll 1$, the LF QPO is at low frequencies, but is
weak and rather broad.  The corresponding energy spectra are
dominated by a hard power law (photon index $\Gamma<2$) which
rolls over at around 100 keV (low/hard spectral state). As the
mass accretion rate increases, the low frequency QPO increases in
strength and coherence as well as frequency, while the power law
spectrum softens and the disc increases in strength relative to
the power law. The QPO is at its highest frequencies, is
strongest and most coherent where the spectrum has both a strong
disc component and a strong soft tail of emission to higher
energies ($\Gamma>2.5$). After this, the LF QPO frequency remains
more or less constant as the tail declines and hardens to $\Gamma
\sim 2.2$, leaving the spectra dominated by the disc component
(high/soft state), though it becomes harder to follow the QPO as
increasing contribution from the stable disc (thermal dominant
state) swamps the signal from the variable tail. (e.g. McClintock
\& Remillard 2006 and references therein)

The only major difference expected in scaling these models up to
the supermassive black holes in AGN is the decrease in disc
temperature down to the UV band. The X-ray spectra of AGN spectra
should then be always dominated by the tail, irrespective of
spectral state (e.g. Done \& Gierli{\'n}ski 2005).  The predicted
change in shape of the tail with spectral state/mass accretion
rate gives an explanation for the variety of spectral properties
seen in {\em unobscured} subtypes AGN such as Broad Line AGN,
Narrow Line Seyfert 1's (NLS1) and LINERs (e.g. Middleton et al.
2008).  The generally hard X-ray spectra seen in LINERS could be
explained as a low $L/L_{\rm Edd}$ flow (e.g. Yuan et al. 2008
but see Maoz 2007), while broad line AGN with a strong UV disc
component and weak X-ray $\Gamma\sim 2$ tail would be analogous
to the high/soft state in BHB. The NLS1 probably have the highest
mass accretion rates, so have lower mass for a given luminosity
(Boroson 2002), and their steeper X-ray spectra (Brandt, Mathur
\& Elvis 1997; Leighly 1999; Shemmer et al. 2006) make them
natural counterparts for the very high state (Pounds, Done \&
Osborne 1995; Murashima et al. 2005; Middleton et al. 2007,
M$^c$Hardy et al. 2007) where all QPOs (both high and low
frequencies) are strongest. Intriguingly, RE J1034+396 is a NLS1,
so its QPO detection is consistent with these scaling models.

All this supports models where the underlying physics of the
accretion flow is very similar between BHB and AGN. However, the
scaling clearly breaks down under more detailed study of the
X-ray spectra from high mass accretion rate objects. These should
have spectra dominated by the steep tail which is typical of the
high and very high states, with no disc emission in the X-ray
band. Yet {\em all} the high mass accretion rate AGN show an
excess below ~1~keV (Gierli{\'n}ski \& Done 2004; Brocksopp et
al. 2006).  This 'soft X-ray excess' is completely inconsistent
with the expected disc component, both as its temperature is
higher than predicted from the mass and mass accretion rate of
these AGN (e.g. Bechtold et al. 1987), and as its shape is much
smoother than a sharply peaked disc spectrum (Czerny et al. 2003;
Gierli{\'n}ski \& Done 2004). Either the soft X-ray excess is an
additional component which breaks the scaling between AGN and
BHB, or it is produced by some external distortion of the
intrinsic emission. Obviously it is very important to distinguish
between these alternatives and the objects with the strongest
soft excesses offer the most stringent constraints.

RE J1034+396 has one of the largest soft excesses known, which
appears to connect smoothly onto the enormous EUV peak of its
spectral energy distribution (Middleton et al. 2007).  The size
and temperature of this peak is extreme even amongst NLS1's
(Puchnarewicz et al. 2002, Casebeer et al. 2006; Middleton et al.
2007), and an obvious question is whether this extreme soft
excess is somehow linked to the detection of the QPO. Here we use
XMM-Newton data on RE J1034+396 to test the various models for
the soft X-ray excess, and speculate on its connection to the
QPO.

\section{The origin of the soft excess in high mass accretion rate AGN}

There are multiple models for the origin of the soft X-ray excess
seen ubiquitously in high mass accretion rate AGN. The most
obvious possibility is that it is related somehow to the disc.
One way to do this is if the mass accretion rate is super
Eddington, so the disc spectrum is distorted by advection of
radiation in the very optically thick flow. Such slim discs
(Abramowicz et al. 1988) have spectra which are less peaked than
a standard disc (Waterai et al. 2000) so give a better fit to the
shape of the soft excess (Mineshige et al. 2000; Wang \& Netzer
2003; Haba et al. 2008). Another way to modify the shape of the
(standard or slim) disc emission is if it is Comptonized by low
temperature electrons (Czerny \& Elvis 1987; Kawaguchi 2003),
perhaps produced by a hotter skin forming over the cooler disc
(Czerny et al. 2003). However, the derived temperature for this
skin is remarkably similar for all objects at 0.1--0.2~keV
despite a large range in black hole mass (and hence disc
temperature: Czerny \etal 2003; Gierli{\'n}ski \& Done 2004,
Crummy \etal 2006). This argues against it being related to the
disc, so it is unlikely to be a true continuum component.

Instead, a constant energy is most easily explained through
atomic processes, in particular the abrupt increase in opacity in
partially ionized material between $\sim$0.7--2 keV due to
O{\small VII}/O{\small VIII} and Fe transitions. This results in
an increase in transmitted flux below 0.7~keV, which could
produce the soft excess either from absorbtion in optically thin
material in the line of sight, or from reflection by optically
thick material out of the line of sight.  In both models the
observed smoothness of the soft excess requires large velocity
smearing (velocity dispersion $\ga$0.3 c) in order to hide the
characteristic {\em sharp} atomic features, but with this
addition then both reflection and absorption models fit the shape
of the soft excess equally well (Fabian \etal 2002, 2004; GD04;
Crummy \etal 2006; Chevallier \etal 2006; Schurch \& Done 2006;
Middleton et al. 2007; Dewangan et al. 2007, D'Ammando et al.
2008). Such high velocities are naturally produced only close to
the black hole, so both absorption and reflection models predict
that the soft excess arises in regions of strong gravity.
However, both models also require some extreme, and probably
unphysical, parameters. In the reflection model, the inferred
smearing can be so large that the required emissivity must be
more strongly centrally peaked than expected from purely
gravitational energy release, perhaps pointing to extraction of
the spin energy itself. The amount of reflection required to
produce the soft excess can also be extreme. Quasi- isotropic
emission sets a limit to size of the soft excess of no more than
a factor of 2--3 above the harder continuum emission (Sobolewska
\& Done 2007), yet the strongest observed soft excesses are a
factor $4$ larger than this (a factor $8-10$ above the extrapolated
continuum), requiring that the intrinsic
illuminating spectrum is strongly suppressed ({\it e.g.}  Fabian
\etal 2002). This issue becomes even more problematic when
incorporating any pressure balance condition (such as hydrostatic
equilibrium) as this strongly limits the section of the disc in
which partially ionized material can exist, hence makes a much
smaller soft excess (Done \& Nayakshin 2007; Malzac, Dumont \&
Mouchet 2005).

Conversely, in the smeared absorption model, pressure balance rather
naturally produces the required partially ionized zone (Chevallier et
al. 2006). However, these models also imply extreme velocities, with a
peak outflow velocity of $\sim$~0.8c required in order to smooth away
the characteristic absorption features from a smooth wind which covers
the source (Schurch \& Done 2007, 2008). This is much larger than the
maximum velocity of $\sim$~0.2-0.4c produced by models of a radiation
driven disc wind (Fukue 1996), so the absorber must instead be
associated with faster material in this description. The obvious
physical component would be a jet/magnetic wind, yet the amount of
material is probably far too large to be associated with either of
these (Schurch \& Done 2007; 2008).

Thus there are physical problems with all of these models: the
Comptonized disc cannot explain the narrow range in inferred
temperatures, smeared reflection cannot produce the strongest
soft excesses seen if the disc is in hydrostatic equilibrium, and
smeared absorption requires a fast wind/jet which is not readily
associated with the properties of any known component.

An alternative possibility for the absorption model is that the
characteristic absorption features are masked by dilution instead
of smearing, perhaps due to the wind becoming clumpy so that it
partially covers the source (Boller et al. 1996; Tanaka et al.
2004; Miller et al. 2007; 2008). These models lack diagnostic
power as the potential complexity of the wind means that any
number of partial absorbers can be added until the model fits the
observed spectra. Nonetheless, this may correspond to the more
messy reality of high Eddington fraction winds from UV luminous
discs (Proga \& Kallman 2004). Some of the clumps could even have
high enough column to produced significant reflected emission as
well (Malzac et al. 2005; Chevallier et al. 2006), giving a
complex mix of reflection and partial absorption from a range of
different columns, velocities and ionization states of the
material.

It is important to distinguish between these very different
potential origins of the soft excess as they make very different
predictions about the environment and geometry of the accreting
material close to the black hole. The extreme broad iron lines
required in the pure reflection models are not required in the
absorption models (either smeared: Sobolewska \& Done 2007; Done
2007 or partial covering: Miller et al. 2008) as the broad
feature redwards of the iron line is fit instead by continuum
curvature. All the absorption models require the presence of
material above the inner disc, probably in the form of a wind,
whereas the reflection models instead suggest a clean line of
sight to the inner disc.  The large velocity shear in the smeared
wind model requires that the material is strongly accelerated, so
it potentially carries an enormous amount of kinetic energy
(Chevallier et al. 2006; Schurch \& Done 2006) with corresponding
impact on AGN feedback/galaxy formation.

Spectral fitting alone cannot distinguish between these very
different spectral models in the 0.3--10 keV bandpass (Crummey et
al. 2007; Sobolewska \& Done 2007; Middleton et al. 2007; Miller
et al. 2007; 2008). Variability gives additional information, but
the smeared reflection and smeared absorption models are known to
be able to predict similar variability patterns (Ponti et al.
2005; Gierli{\'n}ski \& Done 2006). Here for the first time we
fit {\em all} of these models to the data and calculate their
predicted variability to see which description of the soft excess
best matches the {\em simultaneous} constraints from both
spectral and variability data.

\section{Data Extraction}

{\it XMM-Newton} observed RE J1034+396 on 2007-05-31 and
2007-06-01 for about 93 ks (observation id. 0506440101,
revolution no. 1369). We extracted source and background light
curves and noticed background flares and data gaps in the final
~7 ks. Therefore, we excluded this data segment from analysis and
used 84.3 ks of clean data starting on 2007-05-31 20:10:12 UTC
for further analysis.

\subsection{Spectra}

We selected data from the PN (patterns 0--4) and MOS (patterns
0--12) in a region of radius 45 arcsec. The data are very similar
in shape to a previous 16 ks XMM-Newton observation, but are much
higher quality due to the longer exposure.  Because of the
extreme softness of this particular source both MOS and PN were
piled up so we excised the central regions of the image out to
7.5 arcsec radius so as not to be affected by this. Background
was taken from 6 source free regions of the same size.

We use only the MOS data in the 0.3--10~keV range for spectral
fitting as there are well known discrepancies at soft energies
between the PN and MOS spectra which makes simultaneous fitting
of these two instruments very difficult.

We use {\sc xspec} version 11.3.2, and fix the minimum galactic
absorption at $1.31\times10^{20}$cm$^{-2}$ in all fits, but also
allow a separate neutral absorption column to account for
additional absorption in the host galaxy.

\subsection{Energy dependence of the variability: rms spectra}

\begin{figure}
\begin{center}
\begin{tabular}{l}
 \epsfxsize=8cm \epsfbox{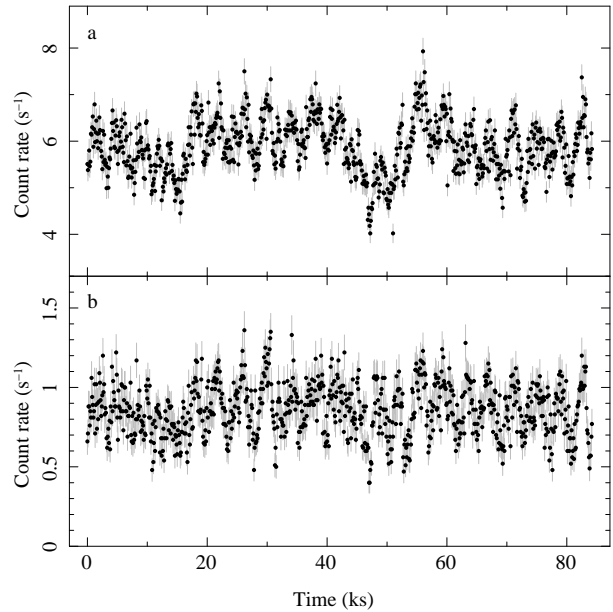}
\end{tabular}
\end{center}

\caption{Light curve of {\it XMM-Newton} (all detectors added
together) observation of RE J1034+396 in (a) 0.3--10 and (b)
1--10 keV band. Clearly the `dip' occurring $\sim$50 ks into the
observation has little effect upon the light curve at high
energies. The starting time of this observation was 2007-05-31
20:10:12 UTC.} \label{fig:lc}
\end{figure}

A successful model must be able to describe {\em both} the
spectrum {\em and} the variability. Fig.~\ref{fig:lc}a shows the
full light curve for these data, with the clear QPO which is
remarkably coherent in the latter part of the observation (25--85
ks: G08). There is also a large scale drop in flux at 40--55 ks.
This looks very similar to an occultation event such as those
recently recognised in other AGN (McKernan \& Yaqoob 1998; Gallo
et al. 2004; Risaliti et al. 2007; Turner et al. 2008).
Fig.~\ref{fig:lc}b shows the light curve in the high energy
bandpass, where this dip is {\em not} present. This shows that
there is clearly energy dependent variability in this event which
is not present in the rest of the light curve. Thus the
variability is {\em complex} and made from more than one
component.

\begin{figure*}
\begin{center}
\begin{tabular}{cc}
 \epsfxsize=8cm \epsfbox{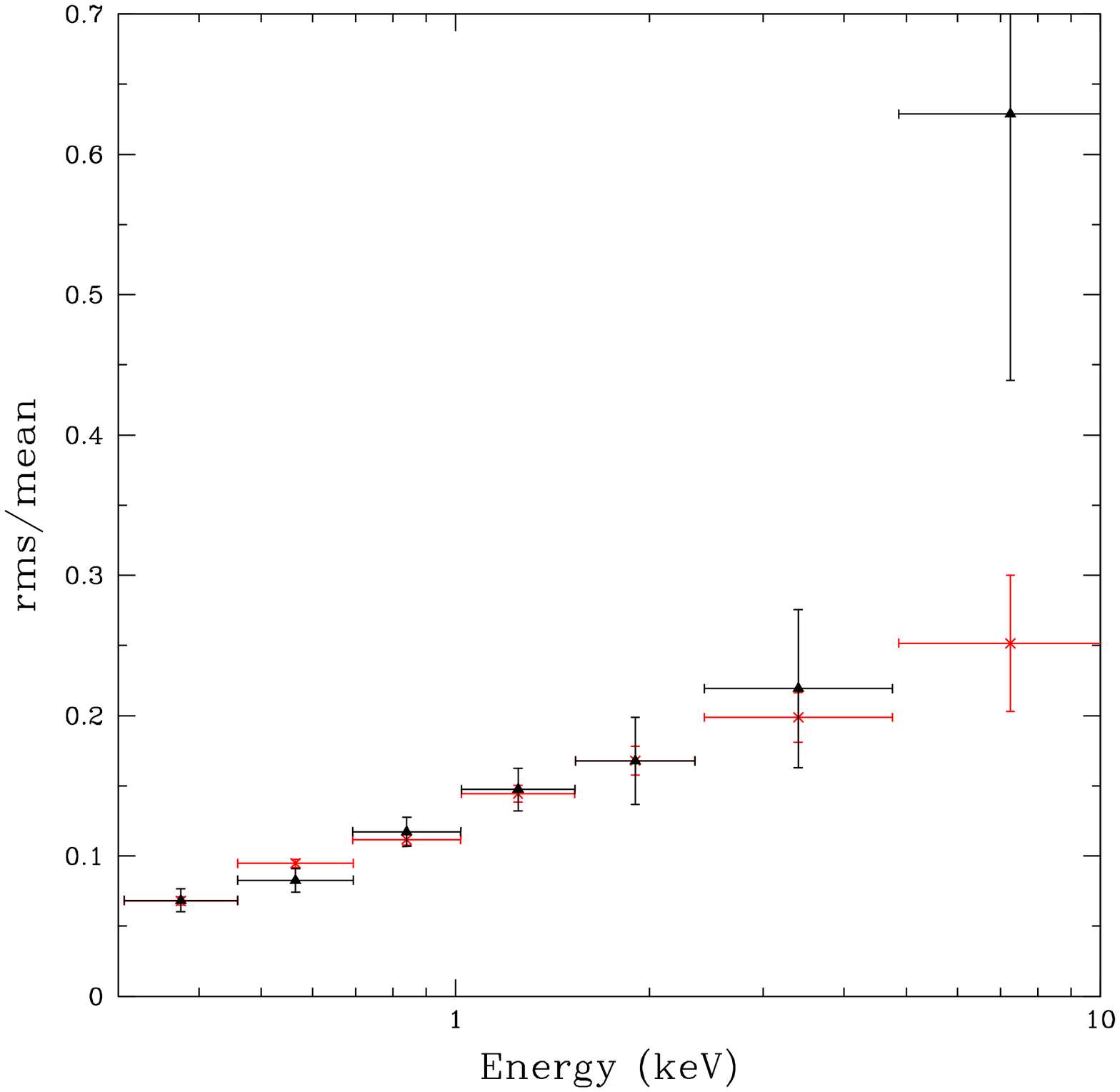} & \epsfxsize=8cm \epsfbox{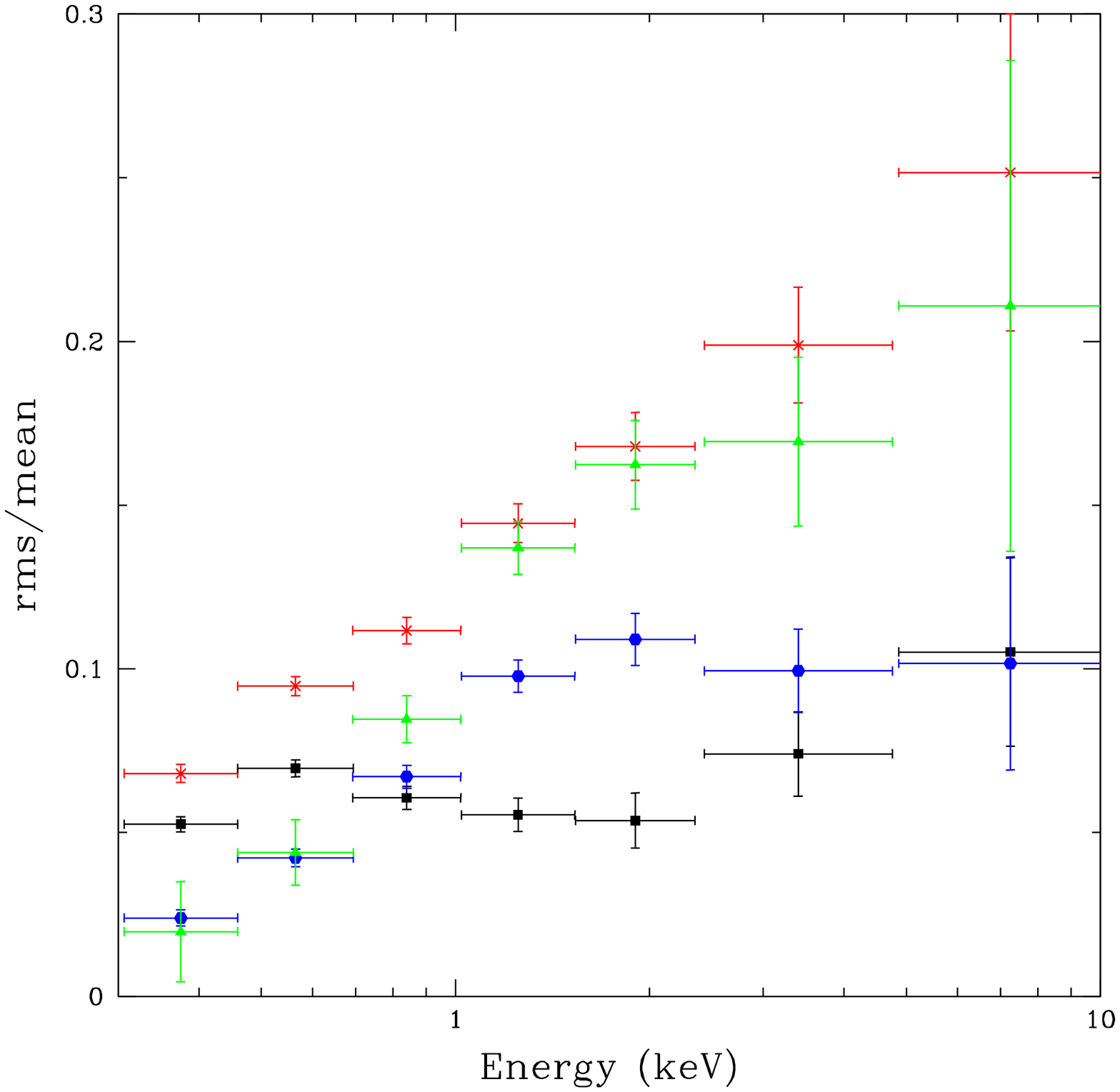}
\end{tabular}
\end{center}

\caption{Rms spectra. Panel (a) shows the rms from the data with
the PSF core excised (black triangles). The rms rises linearly
with energy but the uncertainties at high energies are large due
to the low number of counts. Better statistics at high energy can
be obtained by including all the data (red crosses). Pileup
transfers photons from low energy to high energy, but the lack of
variability in the data at low energy means that this effect
dilutes the high energy rms. Thus these data give a lower limit
on the variability at high energies. Panel (b) shows the
different rms patterns produced from different timescale
variability using all the data. The red crosses show the total
variability as in (a), while the black squares show the long
timescale variability (11.2--5000 $\mu$Hz), the blue circles show
the rms of the QPO alone (frequency of 270 $\mu$Hz) and the green
triangles show the rms of the rapid variability, including the
QPO (135--5000 $\mu$Hz).} \label{fig:rms}

\end{figure*}

We explore this further by calculating the root mean square
fractional variability amplitude (hereafter we will use the term
`rms' for simplicity) as a function of energy (see Edelson et al.
2002; Markowitz, Edelson \& Vaughan 2003 and Vaughan at el.
2003). We first calculate this for the total light curve with
100-s binning. Fig.~\ref{fig:rms}a shows this rms spectrum rising
smoothly with energy. This behaviour is quite unlike the rms
spectra seen from other NLS1's although these can show a variety
of shapes, including flat (e.g. Gallo et al. 2007; O'Neill et al.
2007), flat with a peak at ~2~keV (Vaughan \& Fabian 2004; Gallo
et al. 2004; Gallo et al. 2007), and falling but with a peak at
2~keV (Ponti et al. 2006; Gallo et al. 2007; Larsson et al.
2008).

The rms is dominated by noise above $\sim 4$~keV as the count
rate at high energies is very low due to the steep spectrum. In
order to extend the rms to higher energies we extract light
curves from the full source region, without excising the core to
correct for pileup (red points in Fig.~\ref{fig:rms}b). This
means that some fraction of the hard photons originate from much
lower energies, but the steeply rising rms spectrum means that
the variability of these pileup photons is rather small. Thus
pileup adds an approximately constant offset to the hard
spectrum, and so should dilute the variability seen at high
energies. Instead, the rms spectrum continues to rise in a rather
smooth fashion at high energies (red points in
Fig.~\ref{fig:rms}b), showing that this is a good lower limit to
the total variability at high energies.

The smooth increase of the rms as a function of energy seems
initially most likely to be from a single component. However,
there is clear evidence of different processes contributing to
the variability from the different energy dependence seen in the
dip event (Fig.~\ref{fig:lc}).  We explore this by re-calculating
the rms on a range of timescales by changing the bin time,
$\Delta T$, of the light curves. The rms is the square root of
the integrated power in the light curve from frequencies of 1/$T$
to 1/(2$\Delta T$) Hz, where $T$ = 84300 s is the length of the
light curve. Thus the original binning of 100 s means that the
rms at each energy is the integral of the power spectrum from
11.2 to 5000 $\mu$Hz.  We compare this to the rms of the long
timescale variability by increasing the binning to 3700 s (the
QPO period), i.e. corresponding to the frequency range 11.2--135
$\mu$Hz. This is shown by the black data in Fig.~\ref{fig:rms}b,
and has a very different shape to the total rms (red points),
with a similar amount of variability at low energies, but a sharp
break around 0.7~keV so that the variability at high energies is
much lower. Subtracting this long timescale rms from the total
rms (in quadrature) gives the rms of the rapid variability
(including the QPO at 270 $\mu$Hz), i.e. the integral of the
power spectrum from 135 to 5000 $\mu$Hz (green points in
Fig.~\ref{fig:rms}b). This has a sharp rise with increasing
energy at low energies.  We can compare this directly to the rms
of the QPO by calculating this from the folded light curve (blue
points).  This is indeed very similar in shape and normalization
to the rms of rapid variability.  This is expected as the QPO
forms a large fraction (more than half) of the variability power
in this frequency range.  However, the shape at high energies is
marginally different, with the rapid variability appearing to
continue rising with energy while the QPO saturates at an rms of
$\sim$0.2. Given the potential problems with pileup, and the size
of the uncertainties, this difference is probably not
significant.

\section{Simultaneous constraints from spectra and variability}

In the following sections we take each model for the soft excess
and fit it to the 0.3--10~keV spectrum. The models are listed in
Table \ref{tab:models}. The fit results are shown in Table
\ref{tab:fit_results}. Then we inspect this best-fitting model to
see how varying these components might produce the observed
energy dependence of the variability. We focus here on the rapid
variability, which is dominated by the QPO, as it is clear that
the longer timescale variability may contain contributions from
multiple processes. We calculate the simulated rms spectra by
randomly varying a spectral model parameter with the mean equal
to the best-fitting value and a certain standard deviation
(Gierli{\'n}ski \& Zdziarski 2005). We calculate the rms spectrum
at high energy resolution, but then bin this to the same
resolution as the rms for direct comparison.

\subsection{Separate component for the soft excess: {\sc disk}, {\sc slim} and {\sc comp}}
\label{sec:comp}

\begin{figure}
\begin{center}
\begin{tabular}{c}
\leavevmode
\epsfxsize=8cm \epsfbox{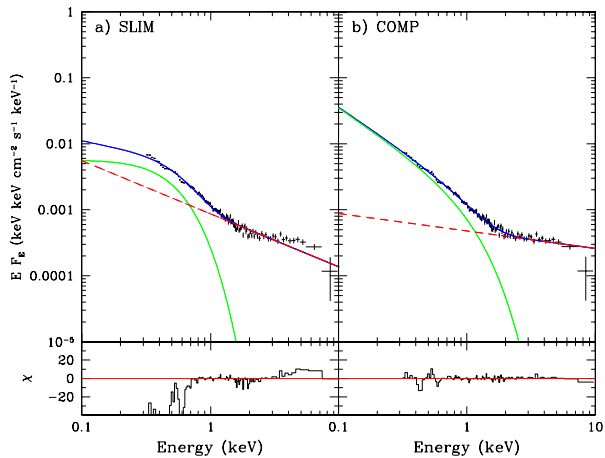}
\end{tabular}
\end{center}

\caption{Spectral models for a separate component for the soft
excess: The upper panel shows the deconvolved spectra, while the
lower panel shows residuals to the fit. (a) uses an advective
disc to describe the soft excess while (b) uses a Comptonized
disc (green solid curve). Both models require a separate power
law to fit the spectrum above 2~keV (red dashed line) to produce
the total spectrum (blue solid line). The lower panel shows the
residuals to the model fit. It is clear that the smooth curvature
of the soft excess rules out the simple advective disc model,
while it is well described by the Comptonized disc.}
\label{fig:disc_panel}
\end{figure}

We first test the models where the soft excess is a separate
emission component. A standard accretion disc spectrum ({\sc
diskbb} in {\sc xspec}) is much more strongly peaked than the
data, giving a very poor fit, $\chi^2_\nu$ = 517/211, showing
that this is an unacceptable description of the broad band
curvature. Instead we test models of an advective (slim) disc,
however the best estimates for mass give a mass accretion rate of
$\sim 0.3 L_{\rm Edd}$, not formally high enough for advection to
be important. Theoretical calculations show that the spectrum of
a slim disc can be approximated by sum of blackbodies, similar to
the standard disc models, but with $T \propto r^{-p}$, where
$0.5<p<0.75$ (the higher limit is the standard disc profile,
while the lower limit is a fully advection dominated disc:
Waterai et al. 2000). Fig.~\ref{fig:disc_panel}a shows that such
models ({\sc diskpbb}, available as an additional local model for
{\sc xspec}) are still not broad enough to match the shape of the
soft excess ($\chi^2_\nu$ = 457/210). Instead, a low temperature
Comptonization component ({\sc comptt}) gives a reasonable fit to
the data, except for the features below $\sim$0.6 keV
(Fig.~\ref{fig:disc_panel}b). Hereafter we refer to this model as
{\sc comp}.

It is plain from the residuals to this fit (lower panel of Fig
3b) that there are weak atomic features at low energies,
especially at $\sim$0.6~keV. Adding a series of narrow Gaussians
gives a reduction in $\chi^2$ of $\sim 30$ for significant
emission features at $\sim 0.55$, $0.74$ and $0.87$~keV. These
are consistent with He--like O Ly$\alpha$ and He- and H-like O
radiative recombination continua i.e. indicating photoionized
material.  However, there are no corresponding narrow lines in
the RGS data, so these features must be intrinsically broad.
Their equivalent width is $\le 10$~eV, so these have negligible
impact on the derived continuum which is the subject of this
paper.

In BHB the power law tail is generically very variable, while the
disc is remarkably constant (Churazov et al. 2000).  If the soft
excess is really related to the disc and is constant, then we
expect the low energy variability to be suppressed, as observed,
because of increasing dilution of the variable power law by the
constant soft excess at low energies.

\begin{figure}
\begin{center}
\begin{tabular}{c}
\leavevmode
\epsfxsize=8cm \epsfbox{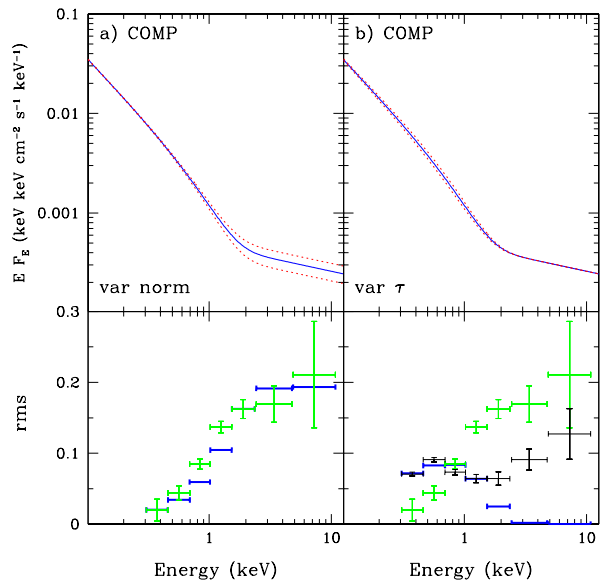}
\end{tabular}
\end{center}

\caption{Variability of a Comptonized disc and power law: The
upper panel shows the mean spectrum (solid blue curve) and $\pm
1\sigma$ variability (dotted red curves). The lower panel shows
the rms produced by this variability (dark thick blue horizontal
lines) compared to that seen in the data on short (light green
crosses). Panel (a) shows the patterns produced by varying the
power law in normalization while keeping the Comptonized disc
constant. The rapid increase in dilution of the power law
variability below 1~keV exactly matches the rapid drop on short
timescales. This is very strong evidence that this spectral
decomposition is correct. (b) shows the pattern produced by
changing the optical depth of the Comptonized disc by 1.5 per
cent, whilst keeping the power law constant. This gives a fairly
good match to the observed long timescale variability (shown as
black crosses in the lower panel).} \label{fig:rms_comptt}
\end{figure}

We quantify this effect by taking the same Comptonization plus
power law model as fits the spectrum of the soft excess, and
varying the normalization of the power law tail by 20 per cent.
This gives a constant rms where the tail dominates (above 2~keV)
and a dramatically increasing suppression of the variability
below this, as the fractional contribution of the constant soft
excess increases (Fig.~\ref{fig:rms_comptt}a). This looks very
similar to the drop observed at low energies in rms of the rapid
variability (and QPO).  This provides very strong evidence that a
two component model is the correct spectral decomposition as both
the spectrum and spectral variability can be easily described in
this model. The soft excess is a separate component above the
power-law tail at low energies, and the rapid variability
(including the QPO) is a modulation of the tail.

The longer timescale variability can also be fit in this model.
The data show a rise in rms with energy to $\sim$1~keV so this
cannot be produced by simply by varying the norm of the
Comptonized disc. Instead, varying the optical depth or
temperature leads to spectral pivoting about the peak of the seed
photons (at $\sim$3$kT_{\rm seed} \approx 0.15$~keV).
Fig.~\ref{fig:rms_comptt}b shows how changing the optical depth
by 1.5 per cent can produce this characteristic rise, which is
then diluted by the constant power law above $\sim 0.8$~keV.
While this does indeed match the broad shape of the long
timescale rms, we caution that there should also be some
contribution to this from partial covering if the dip in the
light curve is indeed an occultation event.

\subsection{Smeared Reflection models: {\sc ref1} and {\sc ref2}}
\label{sec:ref}

We next explore how the reflection model for the soft excess can
fit both the spectrum and spectral variability. We use the
constant density reflection models of Ballantyne, Iwasawa \&
Fabian (2001) (available as a table model in {\sc xspec}). These
models (described in more detail in Ross \& Fabian 2002) become
inaccurate for $\Gamma>2.5$ (Done \& Nayakshin 2007), so instead
we use the spectra tabulated for $\Gamma=2.2$, and multiply the
model by $E^{-(\Gamma - 2.2)}$, where $E$ is energy. This code
also converts the normalization of the reflected emission to that
of the illuminating power law so that the amount of reflection is
parameterized by the solid angle of the illuminated material,
$\Omega/2\pi$ and inclination (Done \& Gierli{\'n}ski 2006),
which we fix at $30\deg$. We smear this using the convolution
version of the {\sc laor} code for the velocity structure of an
extreme Kerr spacetime (Laor 1991).

\begin{figure}
\begin{center}
\begin{tabular}{c}
\leavevmode
\epsfxsize=8cm \epsfbox{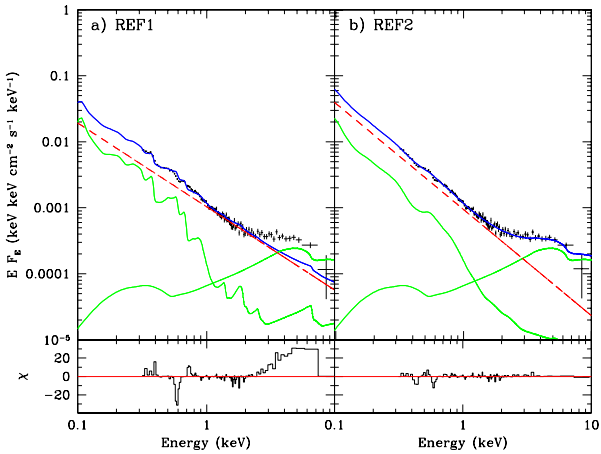}
\end{tabular}
\end{center}

\caption{As for Fig.~\ref{fig:disc_panel} but for a reflection
origin for the soft excess. (a) has a steep power law (red dashed
line) and its ionized, smeared reflection (light green curve)
while (b) has the steep power law illuminating two separate
reflectors, one of which is similarly ionized to that in (a) and
one which is predominantly neutral. The residuals clearly show
that at least two reflectors are required to describe the
spectrum in this model.} \label{fig:refl}
\end{figure}

Fig.~\ref{fig:refl}a shows that this is a very poor fit to the
data ($\chi^2_\nu$ = 612/210), as a single ionization state
reflector cannot simultaneously produce both the hard tail {\em
and} the shape of the soft excess. This is a common feature in
good signal-to-noise data: multiple reflectors are required, one
to fit the soft excess, another to fit the iron line, dispelling
a key attraction of the model (Crummy et al. 2006). Nonetheless,
a double reflector can indeed fit the data well
(Fig.~\ref{fig:refl}b), where the hard tail/iron line region
requires a neutral reflector (modelled using the {\sc thcomp}
code: {\.Z}ycki, Done \& Smith 2000 as the {\sc reflion} code
does not extend down to the very low ionization states required).
We hereafter refer to this model as {\sc ref2}.

This is a very different spectral deconvolution to the one where
the soft excess is a separate Comptonized component ({\sc comp}).
The spectrum at low energies is now modelled by the intrinsic
power law, and instead there is a `hard excess' at high energies
which is formed by dramatically enhanced, strongly smeared,
neutral reflection.

\begin{figure}
\begin{center}
\begin{tabular}{c}
\leavevmode
\epsfxsize=8cm \epsfbox{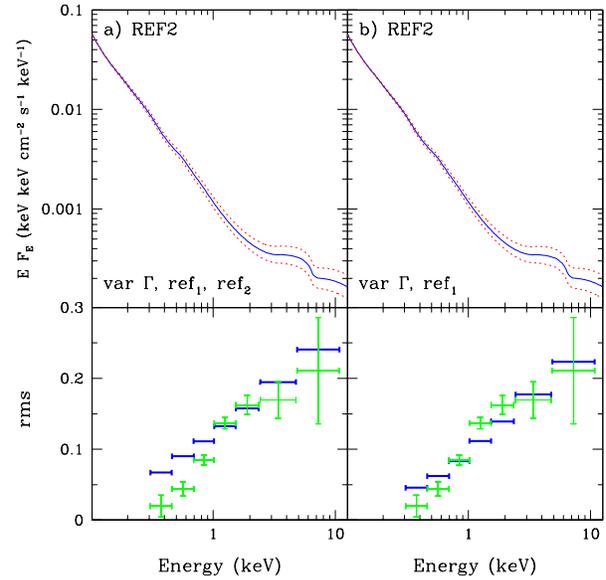}
\end{tabular}
\end{center}
\caption{As for Fig.~\ref{fig:rms_comptt} but for the double reflection
  model for the soft excess. (a) shows the rms produced by pivoting
  the illuminating power law spectral index about the peak in seed
  photon flux at $3kT_{\rm seed}\sim 0.15$~keV when both reflectors
  respond to this illumination. This gives a linear rise
  in rms as a function of energy, which does not match the sharper
  rise around 0.8~keV seen in the rapid variability. (b) shows that
  this can be roughly matched if the ionized reflector which
  contributes mainly at low energies remains constant while the more
  neutral reflector responds to the power law pivoting.}
\label{fig:rms_refl}
\end{figure}

The only way to produce a steeply rising rms is to pivot the
power law about its peak in seed photons (around 0.15~keV, as
before). This gives a linearly rising rms as shown in
Fig.~\ref{fig:rms_refl}a, assuming both reflectors respond to
this changing illumination (though we fix the ionization of each
component for simplicity).

While this would fit the overall rms shown in
Fig.~\ref{fig:rms}b, it plainly cannot match the two very
different shaped components which underlie the variability. The
rapid variability rms shape does {\em not}  rise linearly with
energy, there is a clear inflection point around $\sim$0.8~keV.
This has an obvious origin in the separate Comptonized disc
models discussed above, as this is the energy at which this
component starts to dominate the spectrum and hence dilute the
variability. However, it is possible to match this in the
reflection model by assuming that the ionized reflector remains
constant. Fig.~\ref{fig:rms_refl}b shows how this dilutes the
variability towards lower energies, giving a fairly good fit to
the rapid variability rms.

Constant reflection components are a feature of these models for
the soft excess in other AGN, and may arise from strong light
bending effects (Miniutti \& Fabian 2004). This simultaneously
explains the lack of variability together with enhanced
amplitude and extreme smearing.  However, our spectral
deconvolution and variability require that the constant reflector
is the one which is {\em not} enhanced relative to the
illuminating power law, making this model appear somewhat
contrived in RE~J1034+396.

\subsection{Smeared Absorption Models: {\sc swind} and {\sc xscort}}
\label{sec:swind}

\begin{figure}
\begin{center}
\begin{tabular}{c}
\leavevmode
\epsfxsize=8cm \epsfbox{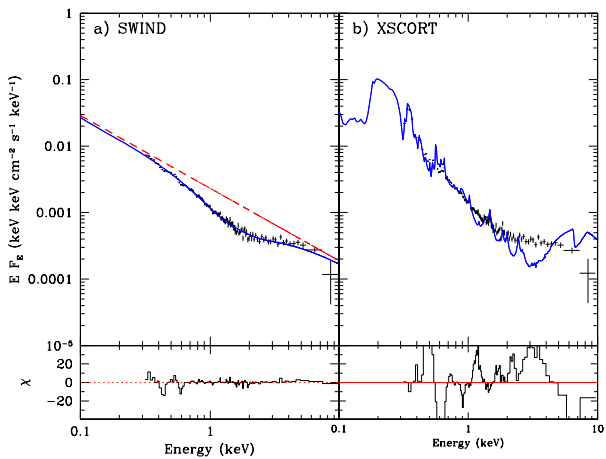}
\end{tabular}
\end{center}

\caption{As for Fig.~\ref{fig:disc_panel} but for a smeared
absorption origin for the soft excess. (a) has a steep power law
(dashed red line) which is distorted by a velocity smeared
absorber which completely covers the source (solid blue curve).
This produces a very good fit to the spectrum, but requires a
large velocity dispersion in the material. (b) shows instead the
results using the expected velocity dispersion from a line driven
disc wind. The velocity spread of 0--0.3c is not sufficient to
smear out the characteristic atomic features.} \label{fig:abs}
\end{figure}

We now fit the smeared absorption models. These assume the source
is completely covered by a partially ionized wind from the disc,
which has a large velocity shear to smear out the characteristic
strong absorbtion lines. We first use the simple {\sc swind}
model (available on the local models page for {\sc xspec}) of
Done \& Gierli{\'n}ski (2004; 2006), where the absorption from a
partially ionized column of material is convolved with a Gaussian
velocity field. This is an excellent fit to the data
(Fig.~\ref{fig:abs}a), but the best fit velocity dispersion goes
to the upper limit of 0.5c allowed in the model.  The gaussian
form of the velocity dispersion means that this corresponds to a
velocity shear greater than the speed of light, which is
obviously unphysical (see Middleton et al. 2007, Schurch \& Done
2007a, b).

\begin{figure}
\begin{center}
\begin{tabular}{c}
\leavevmode
\epsfxsize=8cm \epsfbox{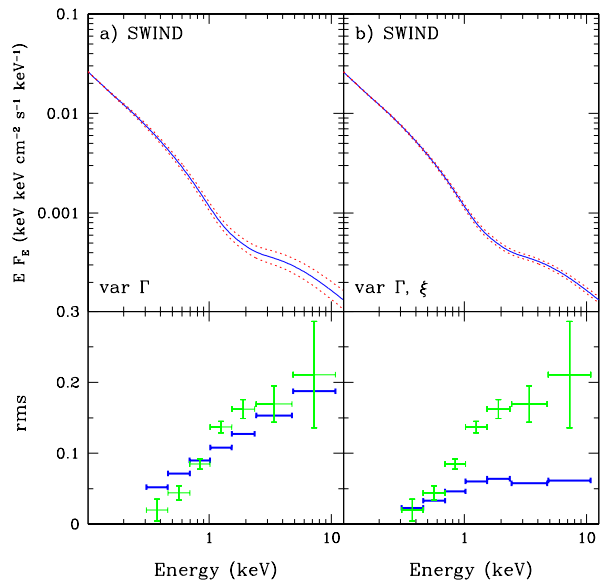}
\end{tabular}
\end{center}

\caption{As for Fig.~\ref{fig:rms_comptt} but for a smeared
absorption model for the soft excess. (a) shows the rms produced
by pivoting the illuminating power law spectral index about the
peak in seed photon flux at $3kT_{\rm seed}\sim 0.15$~keV without
changing the ionization of the absorber. Similarly to
Fig.~\ref{fig:rms_refl}, this linear rise in rms as a function of
energy does not match the sharper rise around 0.8~keV seen in the
rapid variability. (b) shows that this low energy rise can be
matched if the ionization responds to the changes in illumination
as this enhances the variability over 0.7--2~keV where the
absorption dominates. However, this also predicts that the
variability drop above 2~keV, which is not seen in the data.}
\label{fig:rms_abs}
\end{figure}

Instead we fit more physical wind absorption models, calculated
from the {\sc xscort} code, which accelerate the wind from 0 to
0.3c, and use this internal velocity field self-consistently in
the photoionization code.  These models have only been calculated
for a small number of parameters (Schurch \& Done 2007a), so
cannot be properly fit to the data. However, we have taken the
model with parameters closest to those required by the soft
excess spectral shape, and then corrected for the different
spectral index, as in Section \ref{sec:ref}. Fig.~\ref{fig:abs}b
shows that while this gives a strong soft excess, the velocity
shear is insufficient to smooth the strong atomic features in the
0.7--3~keV band into the observed smooth rise, so this gives a
very poor fit to the data ($\chi^2_\nu$ = 2153/212). Thus the
absorption does not arise in a line driven disc wind, but instead
must be connected to much faster moving material in this model.

We explore the variability properties of the fast wind
Fig.~\ref{fig:abs}a, since the more physical wind
(Fig.~\ref{fig:abs}b) does not fit the data.  Because the whole
spectrum is again made from the intrinsically steep power law,
then simply pivoting this at the peak flux of the seed photons
gives a linearly rising rms (Fig.~\ref{fig:rms_abs}a), similar to
that of Fig.~\ref{fig:rms_refl}a. Instead, allowing the
ionization of the fast wind to change in response to this
changing illumination means that the variability is enhanced over
the range where the partially ionized material has most effect on
the spectrum i.e. 0.7--2~keV (Gierli{\'n}ski \& Done 2006).
Fig.~\ref{fig:rms_abs}b shows how this does not produce enough
enhancement of the variability to follow the sharp rise in rms at
$\sim$0.7~keV and it also predicts that this enhancement does not
affect the spectrum above $\sim$2~keV so the rms at high energies
is also too small.

Thus while the smeared absorption model is a good fit to the spectrum,
it cannot easily match the variability patterns seen, making it a less
attractive solution.

\subsection{Partial covering models: {\sc ionpcf1} and {\sc ionpcf2}}
\label{sec:ionpcf}

\begin{figure}
\begin{center}
\begin{tabular}{c}
\leavevmode
\epsfxsize=8cm \epsfbox{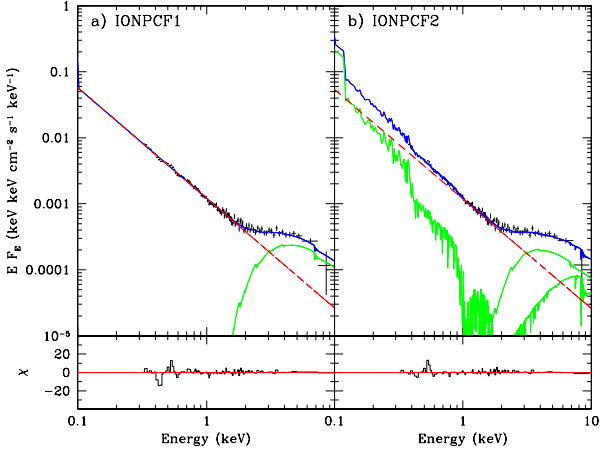}
\end{tabular}
\end{center}

\caption{As for Fig.~\ref{fig:disc_panel} but for a partial
covering model for the soft excess. (a) has an intrinsically
steep power law, part of which is strongly absorbed by nearly
neutral material (light green curve) and part of which is
unabsorbed (dashed red line). (b) shows a marginally better fit
to the spectrum, where there is a second absorbed component from
material which is ionized, so it also contributes at low
energies. } \label{fig:pcf}
\end{figure}

If instead the absorption is clumpy then it can lead to clouds
which partially cover the source. These can be ionized as they
are illuminated by the central source, so we model this using
partial covering by partially ionized material ({\sc zxipcf},
available as an additional model for {\sc xspec}).  This fits the
data fairly well (Fig.~\ref{fig:pcf}a) for a column with very low
ionization covering 75 per cent of the source. We call this model
{\sc ionpcf1}. A second partially ionized, partial covering
component is marginally significant ($\Delta\chi^2=12$ for 4
additional parameters), but this must be outflowing, with a
blueshift of $0.12\pm 0.01c$. This is similar to the velocities
sometimes seen in highly ionized K$\alpha$ absorption lines from
iron in high mass accretion rate AGN (e.g. the compilation Reeves
et al. 2008), and also close to the theoretical expectations of
the maximum velocity of a UV line driven disc wind (Proga et al.
2004). Hereafter we call this model {\sc ionpcf2}.

\begin{figure}
\begin{center}
\begin{tabular}{c}
\leavevmode
\epsfxsize=8cm \epsfbox{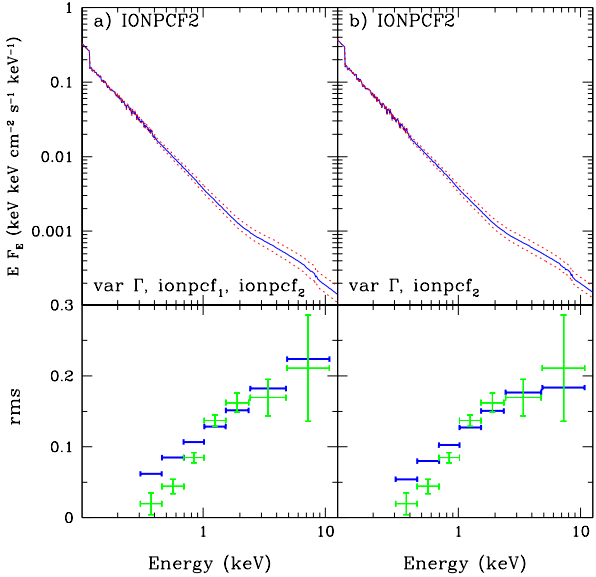}
\end{tabular}
\end{center}
\caption{As for Fig.~\ref{fig:rms_comptt} but for the partial covering
  model for the soft excess.
  (a) shows the rms produced by pivoting the
  illuminating power law spectral index about the peak in seed photon
  flux at $3kT_{\rm seed}\sim 0.15$~keV without changing the ionization of
  the absorber. Similarly to Fig.~\ref{fig:rms_refl}a
  and Fig.~\ref{fig:rms_abs}a, this linear rise
  in rms as a function of energy does not match the sharper
  rise around 0.8~keV seen in the rapid variability. (b) shows that
  this low energy rise is better fit if the ionized absorber which
  contributes at low energies is held constant while the power law
  pivots, but the model's predicted rise at $\sim$0.8~keV is still stronger than is observed.}
\label{fig:rms_ion}
\end{figure}

Similarly to Fig.~\ref{fig:rms_refl}a and
Fig.~\ref{fig:rms_abs}a, pivoting the power law at the peak flux
of seed photons gives a linearly rising rms
(Fig.~\ref{fig:rms_ion}a) which does not match the sharper rise
in rms seen in the data. Again, dilution by a constant component
at low energies is required, so Fig.~\ref{fig:rms_ion}b shows the
variability pattern produced by assuming that the ionized
absorption component at low energies remains constant while the
rest of the spectrum varies as before. Physically this could
arise from light travel time delays if this is scattered from
ionized material at some distance from the source. This produces
some suppression of variability at low energies, but not as much
as required to fit the rms of the rapid variability.

\begin{table*}
\begin{tabular}{lll}
\hline
Name & Description & {\sc xspec} syntax \\
\hline
{\sc disk}     & Standard disc & {\sc wabs(diskbb+powerlaw)}\\
{\sc slim}     & Advective disc & {\sc wabs(diskpbb+powerlaw)} \\
{\sc comp}     & Comptonized component & {\sc wabs(comptt+powerlaw)} \\
{\sc swind}    & Smeared wind absorption & {\sc wabs*swind(powerlaw)} \\
{\sc xscort}   & Ionized absorption/emission & {\sc wabs(xscort)} \\
{\sc ref1}     & Ionized reflection (single) & {\sc wabs(powerlaw+conline*reflbal(powerlaw))} \\
{\sc ref2}     & Ionized reflection (double) & {\sc wabs(powerlaw+conline(thcomp)+conline*reflbal(powerlaw))} \\
{\sc ionpcf1}  & Single ionized partial covering & {\sc zxipcf*wabs(powerlaw)} \\
{\sc ionpcf2}  & Double ionized partial covering & {\sc zxipcf*zxipcf*wabs(powerlaw)} \\
\hline
\end{tabular}
\caption{Summary of the models used in the paper.}
\label{tab:models}
\end{table*}

\begin{table}
\begin{tabular}{llll}
\hline
Model & Parameter & Value & $\chi_\nu^{2}$/d.o.f   \\
\hline

{\sc comp}    &  $\Gamma$                       & $2.28^{+0.13}_{-0.10}$  & 289/210\\
              & $N_{\rm PL}$ ($\times10^{-4}$) & $4.9^{+0.8}_{-0.3}$ \\
              & $kT_e$ (keV)                   & $0.26\pm0.03$ \\
              & $N_{\rm comptt}$               & $10.6^{+1.2}_{-0.8}$ \\
\hline
{\sc swind}   & $\Gamma$                       & $3.10^{+0.02}_{-0.05}$ & 306/210\\
              & $N_{\rm PL}$ ($\times10^{-4}$) & $23.1^{+0.9}_{-0.7}$ \\
              & lg $\xi$                       & $2.82^{+0.09}_{-0.02}$ \\
              & $\sigma$                       & $0.50^{+**}_{-0.05}$ \\
\hline
{\sc ref2}    & $\Gamma$                       & $3.61^{+0.08}_{-0.04}$ & 268/206\\
              & $N_{\rm PL}$ ($\times10^{-4}$) & $9.4^{+0.3}_{-2.4}$  \\
              & $\Omega_1/2\pi$                & $0.43^{+1.05}_{0.10}$\\
              & $R_{\rm in,1}$                 & $2.1^{+1.0}_{-0.8}$   \\
              & lg $\xi$                       & $3.02^{+0.16}_{-0.07}$  \\
              & $\Omega_2/2\pi$                & $56^{+15}_{-18}$ \\
              & $R_{\rm in,2}$                 & $3.2^{+0.4}_{-1}$ \\
\hline
{\sc ionpcf1} & $\Gamma$                       & $3.69\pm0.05$ & 287/210\\
              & $N_{\rm PL}$ ($\times10^{-4}$) & $74\pm13$ \\
              & lg $\xi$                       & $0.8^{+0.3}_{-0.5}$ \\
              & $N_H$ ($10^{22}$ cm$^{-2}$)    & $14^{+2}_{-3}$ \\
              & $f$                            & $0.84^{+0.02}_{-0.04}$ \\
\hline
{\sc ionpcf2} & $\Gamma$                       & $3.68^{+0.09}_{-0.13}$ & 233/206\\
              & $N_{\rm PL}$ ($\times10^{-4}$) & $165^{+367}_{-69}$  \\
              & lg $\xi_1$                     & $0.7^{+0.4}_{-0.9}$ \\
              & $N_{H,1}$ ($10^{22}$ cm$^{-2}$) & $12^{+2}_{-5}$ \\
              & $f_1$                          & $0.80\pm 0.06$ \\
              & lg $\xi_2$                     & $2.72^{+0.05}_{-0.5}$ \\
              & $N_{H,2}$ ($10^{22}$ cm$^{-2}$) & $153^{+13}_{-14}$ \\
              & $f_2$               & $0.63^{+0.07}_{-0.09}$ \\
              & $v_2/c$               &  $0.12\pm 0.01$ \\

\hline

\end{tabular}
\caption{Best-fitting parameters of selected spectral models.
Models described in the text that do not fit the spectrum are not
shown here. Error bars are 90 per cent confidence, and ** denotes
a parameter that reached its limit. $\Gamma$ is the photon
spectral index, $N_{\rm PL}$ is power-law normalization at 1 keV,
$\xi$ is the ionization (in the units of erg cm s$^{-1}$),
$\sigma$ is the velocity dispersion in the units of $c$, $\Omega$
is the reflector solid angle, $R_{\rm in}$ is the inner disc
radius (in the units of $GM/c^2$), $f$ is the covering factor.}
\label{tab:fit_results}
\end{table}

\section{Constraints from spectra and variability}

Any viable model for the soft excess must simultaneously fit both
the spectrum and spectral variability with the same model
parameters.  The spectra alone can be fit with a variety of
continua. There is clear curvature but this is degenerate. The
spectra can either be fit with low energy spectral curvature,
together with a ($\Gamma\sim 2.3$) power law at high energies, or
with a steep power law at low energies ($\Gamma\sim 3.6$)
together with curvature at high energies from emerging reflected
({\sc ref2}) or absorbed components ({\sc ionpcf2}), or by a
somewhat less steep power law ($\Gamma\sim 3.3$) with curvature
around 0.7--2~keV from absorption in a rapidly accelerating wind
({\sc swind}). However, the energy dependence of the rapid X-ray
variability (which is dominated by the QPO in these data) breaks
these degeneracies.  The rms spectrum over the whole observation
shows a smooth rise with energy, initially appearing to favour
models where there is a single variable component (pivoting power
law) which forms the spectrum. However, this rms is made up from
two quite dissimilar variability patterns for the short and long
timescales.  Both these contain clear changes in the rms at $\sim
0.7$~keV, with the rapid variability amplitude strongly
increasing with energy at this point while the longer timescale
variability strongly decreases. These combine together in such a
way as to produce an apparently featureless overall rms.

The rms of the rapid variability clearly supports the spectral
decomposition where the soft excess is a separate, low
temperature Comptonization component ({\sc comp}). The
variability drops just where the low temperature Compton
component starts to dominate, so the rms shape is easily produced
from models where the soft excess remains constant and simply
dilutes the variability of a high energy power law tail.  This is
very similar to the sorts of spectral decompositions used in
binary systems, where the QPO (and all other rapid variability)
is associated with the tail, not with the disc.

By contrast, the alternative spectral models (where the soft
excess is an artifact of some distortion on an intrinsic steep
power law spectrum) have much more difficulty in matching the
energy dependence of the rapid variability. Pivoting the steep
intrinsic power law at the peak energy of the seed photons gives
a linearly rising rms, so some additional assumptions are
required to produce the observed faster drop in variability at
low energies.  Changing the ionization of the absorber in the
smeared wind model ({\sc swind}) gives enhanced variability in
the 0.7--2~keV region which matches well to the shape of the rms
at low energies. However, this then drops above 2~keV where
ionization changes no longer affect the spectrum, which does not
match the observed rms.

By contrast, both the double reflector ({\sc ref2}) and double
partial covering model ({\sc ionpcf2}) have ionized components at
low energies, so holding these constant while the rest of the
spectrum responds to the pivoting power law dilutes the low
energy variability. However, physically this seems somewhat
contrived, and both these (but especially {\sc ionpcf2}) have
difficulties in suppressing enough variability at the lowest
energies.

\begin{figure}
\begin{center}
\begin{tabular}{c}
\leavevmode
\epsfxsize=6cm \epsfbox{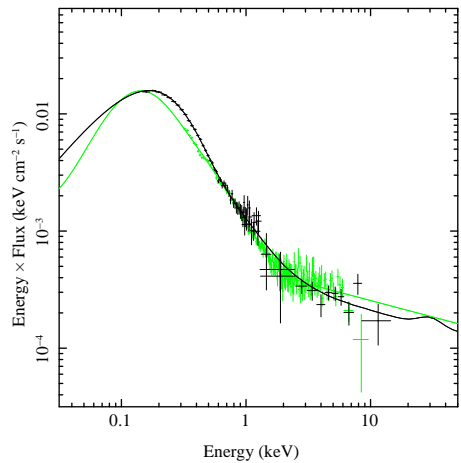}
\end{tabular}
\end{center}

\caption{Comparison of RE~J1034+396 (light green) and GRS
1915+105 (black) energy spectra. Both spectra are unfolded and
unabsorbed using the best-fitting model. The GRS 1915+105
spectrum was shifted in energy by factor 20 and renormalized to
match the RE~J1034+396 spectrum.} \label{fig:1915}

\end{figure}

Thus the combined spectral and spectral variability constraints
strongly favour the model where the soft excess is a true
continuum component, described by a low temperature Comptonized
disc component ({\sc comp}).  The long timescale variability can
also be explained in this spectral decomposition by changing the
temperature and/or optical depth of the Comptonization to give a
predicted rms which matches the data very well. However, the
light curve shows that there can also be discrete energy
dependent events (the dip), which seems more likely to be from an
occultation. If so, then there should also be some contribution
from absorption in the spectrum, and this might also shape the
longer timescale variability.

\section{Constraints from the broad band spectral energy distribution}

This decomposition of the X-ray spectrum into a Comptonized disc with
separate tail to high energies ({\sc comp}) also looks sensible in the
light of the overall spectral energy distribution.  Both the
UV/EUV/soft X-ray shape and the rms variability of the X-ray emission
show that this peaks at $\sim 0.15$~keV (Puchnarewicz et al. 2000;
Casebeer et al. 2006; Fig.~\ref{fig:rms}). Integrating the {\sc comp}
model down to this energy gives a luminosity of $\sim 7\times 10^{43}$
ergs s$^{-1}$.  This is already more than 10 per cent of the
bolometric luminosity of $\sim 5\times 10^{44}$ ergs
s$^{-1}$. However, in the reflection and partial covering models, {\sc
ref2} and {\sc ionpcf2}, the observed soft X-ray flux is only a small
fraction of the total illuminating power law.  The hidden emission
required by both these models is a factor $\sim 10$ larger. This
luminosity must be reprocessed, yet is more than the total bolometric
luminosity of the source!  Thus these two models fail on energetics as
well as requiring somewhat contrived assumptions in order to match the
rms variability and requiring an uncomfortably steep ($\Gamma\sim
3.6$) intrinsic continuum.

This all strongly supports the conclusion that in RE~J1034+396
the soft excess is a Comptonized disc component, connecting
smoothly onto the EUV peak of the spectral energy distribution
and extending out to $\sim 1$~keV. This does not have an obvious
counterpart in the typical black hole binary systems. These can
show convincingly clean disc spectra (thermal dominant state,
high/soft state, ultrasoft state), or higher temperature
Comptonization (10--20~keV: very high state, intermediate state,
steep power law state), but generally they do not show low
temperature Comptonization (see e.g. Done, Gierli{\'n}ski \&
Kubota 2007). However, most of these systems have $L/L_{\rm
Edd}<1$, so they may not be a good guide to the spectra of super
Eddington flows.  The disc structure should change at such high
mass accretion rates, with strong winds potentially increasing
the amount of material in the corona, leading to mass loading of
the acceleration mechanism and decreasing energy per particle.
Whatever the origin, a similar process of low temperature disc
Comptonization probably happens in the unique galactic black hole
binary GRS~1915+105. This is the only black hole binary in our
Galaxy which consistently shows super Eddington luminosities
(Done, Wardzi{\'n}ski \& Gierli{\'n}ski 2004), and it can show
spectra which are dominated by a huge, low temperature
Comptonized disc component, with a weak power law tail to higher
energies. Fig.~\ref{fig:1915} shows how one of these spectra from
GRS~1915+105 (the low $\omega$ state in fig. 7 of Zdziarski et
al. 2005) fits almost exactly onto the XMM-Newton spectrum of RE
J1034+396 with a shift in energy scale by a factor 20. This gives
a mass estimate for RE J1034+396 of $\sim 2\times 10^6 $M$_\odot$
from scaling the disc temperature as $M^{-1/4}$, which seems
quite reasonable (Puchnarewicz et al. 2002).

Understanding the origin of the soft excess in RE~J1034+396 may
not necessarily solve the problem of the soft excess in general.
The spectral energy distribution of this object is plainly rather
different from that in most other high mass accretion rate AGN in
that the soft excess contains a large fraction of the bolometric
luminosity of the source (Middleton et al. 2007). Spectral
distortion models (reflection or absorption) are clearly more
likely to explain a feature carrying only a small fraction of the
bolometric luminosity. Instead, the continuous EUV/soft X-ray
excess in RE~J1034+396 clearly favours a common origin for the
disc and soft excess, unlike that for most other quasars, where
the UV and soft X--rays do not appear to smoothly connect to each
other in individual objects (Haro-Corzo et al. 2007).

\section{Conclusions}

The combined constraints from the spectrum and variability show
the soft excess in RE J1034+396 is most likely a smooth extension
of the accretion disc peak in the EUV probably arising from low
temperature Comptonization of the disc.  This remains more or
less constant on short timescales, diluting the QPO and rapid
variability seen in the power law tail at the low energies where
the soft excess dominates. As in the black hole binary systems,
the QPO is a feature of the tail {\em not} of the disc.

The conclusion that the soft excess is a low temperature
Comptonization of the disc emission may not necessarily be more
widely applicable to other NLS1.  The spectrum of RE~J1034+396 is
unique, so extrapolating results from this object may not be
justified.  In particular, if this object has a very high (super
Eddington?) mass accretion rate then it could enter a new
accretion state (perhaps analogous to the unique galactic binary
GRS 1915+105 and the Ultra Luminous X-ray sources: Roberts 2008)
which may also be the trigger for its so far unique QPO. A
different origin for the soft excess in RE~J1034+396 has the
advantage that it would not require some unknown physical
mechanism to restrict the temperature of the continuum in all
objects to a narrow range (Gierli{\'n}ski \& Done 2004), though
does require a coincidence that this temperature in RE~J1034+396
is so close to that ubiquitously seen for the soft excess.
Nonetheless, it is clearly possible that in selecting the most
extreme soft excess object, we have also selected the one where
the the soft excess has a different origin.

\label{lastpage}

\end{document}